# CRIICAL REGIMES OF INTERNAL GRAVITY WAVE GENERATION

Vitaly V. Bulatov (1), Yuriy V. Vladimirov (1), Vasily A. Vakorin (2)

(1) Institute for Problems in Mechanics, Russian Academy of Sciences, Moscow, Russia

(2) University of Saskatchewan, Saskatchewan, Canada.

**Abstract**

The problem of constructing an asymptotic representation of the solution of the internal gravity wave field exited by a source moving at a velocity close to the maximum group velocity of the individual wave mode is considered. For the critical regimes of individual mode generation the asymptotic representation of the solution obtained is expressed in terms of a zero-order Macdonald function. The results of numerical calculations based on the exact and asymptotic formulas are given.

The investigation of the critical regimes of generation and propagation of internal gravity waves is important for describing vertical and horizontal transfers in natural stratified media, since, by virtue of the relatively small dissipation of the internal waves, when the waves propagate over large distances it is possible to investigate the characteristics of various sources of generation [1]. For example, one of these regimes may be the excitation of waves by a stratified fluid flow with a velocity close to the maximum group velocities of the corresponding modes generated by this internal wave flow. This generation pattern can be observed under certain conditions, including the impingement of a stratified flow on an oceanic shelf. In this case in some spatial zones of the shelf the flow velocity may coincide with the local group velocity for a given bottom depth [3]. By means of numerical calculations it was shown that in the case of unsteady motion of a source in a stratified fluid flow the maximum-amplitude waves are generated at near-critical velocities [2]. The problem considered is linear and, therefore, in what follows, obtaining asymptotic representations of the solutions describing the critical regimes of internal gravity wave generation by point sources also makes it possible effectively to investigate the analogous regimes for arbitrary nonlocal perturbation sources [1].

The elevation η of the field of the internal gravity waves excited by a point mass source of unit intensity which begins to move at the moment $t = 0$ in a stratified fluid layer $-H < z < 0$ can be determined from the problem [1]:

$$L_\eta = \theta(t) \frac{\partial^2}{\partial t \, \partial z_0} (\delta(x - x_0(t)) \, \delta(y - y_0(t)) \, \delta(z - z_0(t))) \qquad (1)$$

$$L = \frac{\partial^2}{\partial t^2} \left( \frac{\partial^2}{\partial x^2} + \frac{\partial^2}{\partial y^2} + \frac{\partial^2}{\partial z^2} \right) + N^2(z) \left( \frac{\partial^2}{\partial x^2} + \frac{\partial^2}{\partial y^2} \right)$$

Where $N(z)$ is the Brunt-Väisälä frequency, $\theta(t) = 0$ when $t < 0$, $\theta(t) = 1$ when $t > 0$, and $(x_0(t), y_0(t), z_0(t))$ is the trajectory of the source. As the boundary conditions we use the "rigid lid" approximation:

$$z = 0, -H : \quad \eta = 0 \qquad (2)$$

The solution of (1), (2) has the form:

$$\eta = \sum_{n=1}^{\infty} \eta_n$$

$$\eta_n = \frac{1}{2\pi} \mathrm{Re} \int_0^\infty \frac{\omega_n^2(k)}{k} \varphi_n(z,k) \int_0^t \frac{\partial \varphi_n(z_0(\tau), k)}{\partial z_0(\tau)} \exp(i\omega_n(k)(t-\tau)) J_0(kr(\tau)) \, d\tau \, dk$$

$$r(\tau) = [(x - x_0(\tau))^2 + (y - y_0(\tau))^2]^{1/2}$$

where $\omega_n(k)$ and $\varphi_n(z,k)$ are the eigenfunctions and the eigenvalues of the corresponding vertical spectral problem [1], and $J_0$ is a zero-order Bessel function.

In what follows, we will consider the rectilinear uniform motion of a source at a constant dept $z_0 = \mathrm{const}$, $y_0 = 0$, and $x_0 = -V\tau$ and an individual mode whose index will be omitted. Using the change of variables $x + V\tau = \xi$ and $x + Vt = \xi_t$, for η we obtain the following expression:

$$\eta = \frac{1}{2\pi V} \mathrm{Re} \int_0^\infty \frac{\omega^2(k)}{k} \varphi(z,k) \frac{\partial \varphi(z_0,k)}{\partial z_0} \int_x^{\xi_t} \exp(i\omega(k)(\xi_t - \xi)) J_0(k\sqrt{y^2 + \xi^2}) \, d\xi \, dk \qquad (3)$$

The steady-state regime is the limit of the solution (3) as $x \to -\infty$ for fixed values of $\xi_t$. Then, considering the solution for the steady-state regime on the traverse of the source $\xi_t = 0$ and using the following relation [4]

$$\int_0^\infty \cos\left(\frac{\omega(k)\xi}{V}\right) J_0(k\sqrt{y^2 + \xi^2}) d\xi = \begin{cases} \dfrac{\cos(y\lambda(k))}{\lambda(k)}, & k > \mu(k) \\ 0, & k < \mu(k) \end{cases}$$

$$\lambda(k) = (k^2 - \omega^2(k)V^{-2})^{1/2}, \qquad \mu^2(k) = \omega^2(k)V^{-2}$$

we obtain the expressions for $\eta$ in the form:

$$\eta = \operatorname{Re} \int_k^\infty \frac{\omega^2(k)}{k} \frac{\exp(iy\lambda(k))}{\lambda(k)} \Phi(k, z, z_0) dk$$

$$\Phi(k, z, z_0) = \frac{1}{2\pi V} \varphi(z, k) \frac{\partial \varphi(z_0, k)}{\partial z_0}$$

$$c = \partial \omega(k)/\partial k \big|_{k=0}$$

where $c$ is the maximum group velocity of the corresponding mode, $K$ is a root of the equation $k^2 V^2 = \omega^2(k)$ when $V < c$, and $K = 0$ when $V > c$.

In what follows, for the sake of simplicity, we will consider the case of an exponentially stratified fluid $N(z) = \text{const}$. We than have

$$\Phi(k, z, z_0) = \frac{1}{VN^2 H^2} \sin\left(\frac{\pi z}{H}\right) \cos\left(\frac{\pi z_0}{H}\right) \equiv A$$

$$K = \frac{\pi}{H}(c^2 V^{-2} - 1)^{1/2} \equiv \varepsilon, \qquad c = \frac{NH}{\pi}$$

$$\eta = A \operatorname{Re} \int_0^\infty \frac{N^2 \exp(i \, ykT^+(k))}{S^+(k)} dk, \quad M > 1 \qquad (4)$$

$$\eta = A \operatorname{Re} \int_0^\infty \frac{N^2 \exp(i \, ykT^-(k))}{S^-(k)} dk, \quad M < 1$$

$$T^\pm(k) = (k^2 \pm \varepsilon^2)^{1/2}(k^2 + b^2)^{-1/2}$$

$$S^{\pm}(k) = (k^2 \pm \varepsilon^2)^{1/2}(k^2 + b^2)^{1/2}$$

$$b^2(1 - M^{-2}) = \begin{cases} \varepsilon^2, & M > 1 \\ -\varepsilon^2, & M < 1 \end{cases}$$

where $b = \pi/H$ and $M = V/c$ is the Mach number of the corresponding mode, and $\varepsilon$ is a parameter describing the deviation of the source velocity $V$ from $c$. Clearly, for small values of $\varepsilon$ the behavior of integrals (4) is determined by small values of $k$

$$\eta \approx A \operatorname{Re} \int_0^\infty c^2 \exp(i\, yk(k^2 + \varepsilon^2)^{1/2} b^{-1})(k^2 + \varepsilon^2)^{-1/2}\, dk, \qquad M > 1$$

$$\eta \approx A \operatorname{Re} \int_0^\infty c^2 \exp(i\, yk(k^2 - \varepsilon^2)^{1/2} b^{-1})(k^2 - \varepsilon^2)^{-1/2}\, dk, \qquad M < 1$$

Using the changes of variable $k = \varepsilon \sinh t$ when $M > 1$ and $k = \varepsilon \cosh t$ when $M < 1$, for any $V$ we can obtain the following expression [4]:

$$\eta \approx A \operatorname{Re} \int_0^\infty c^2 \exp(i\, y\, \varepsilon^2 \sinh(2t) b^{-1})\, dt \qquad (5)$$

$$\eta \approx \frac{A N^2 H}{2\pi} K_0\left(\frac{\pi y \varepsilon^2}{H}\right)$$

where $K_0$ is a zero-order Macdonald function. The expression obtained describes the asymptotics of an individual mode of the internal wave field generated by a source moving at a velocity close to the maximum group velocity of the corresponding mode.

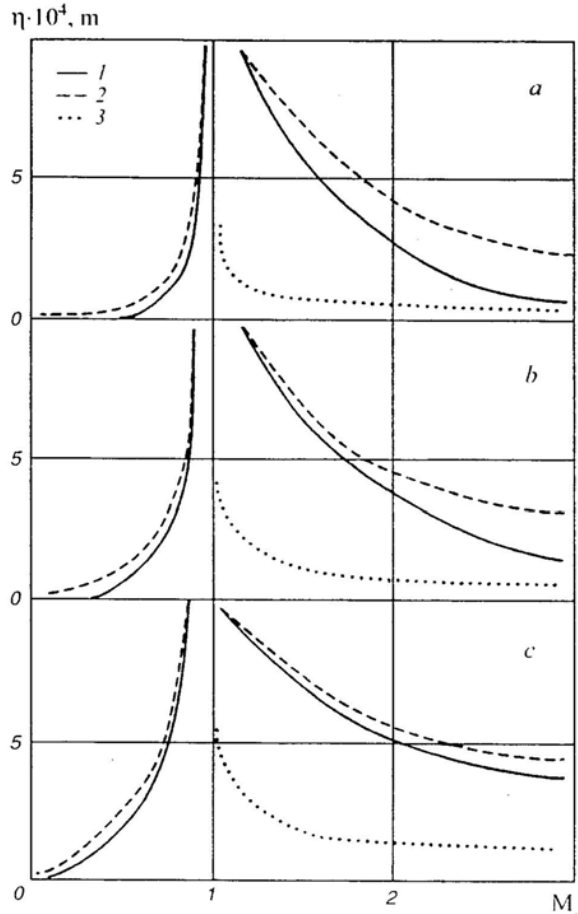

Fig. 1. The first mode of internal gravity wave field elevation: $y/H = 3, 2,$ and $1$ ($a$, $b$, and $c$, respectively).

In order to illustrate the result obtained in Fig. 1 we have reproduced the results of numerical calculations based on the exact formulas (4) (curve *1*) and asymptotic formulas (5) (curve *2*) for $y/H = 1, 2, 3$. All the calculations were carried out for the first mode. As can be seen from these numerical results, the asymptotic representation of the solution obtained correctly describes the exact solution for an individual mode over a fairly wise range of source velocities on the interval from $c/2$ to $2c$.

The asymptotic representation of the solution obtained in the form (5) has a logarithmic singularity at $y = 0$ which is integrable. Clearly, only fields excited by nonlocal sources are physically meaningful; therefore, in order to calculate the wave field induced by a nonlocal source it is sufficient to integrate either the exact or the asymptotic representation of the solution for a point source with the corresponding weight [1].

As might be expected, in accordance with [1], when both $M \gg 1$ and $y/H$ is large (far field), the asymptotic representation of the individual modes of the internal waves should be described by the asymptotic representation expressed in terms of the Airy function as follows:

$$\eta \approx \frac{AN^2}{2b\varepsilon(3\beta yp)^{1/3}} \text{Ai}(y(3\beta yp)^{-1/3}) \qquad (6)$$

$$p = (M^2 - 1)^{1/2}, \qquad \beta = V^4 \alpha (V^2 - c^2)^{-5/2}$$

where Ai is the Airy function and $c$ and $\alpha$ are the first two coefficients of the expansion of the dispersion curve of the corresponding mode $\omega(k)$ at zero $\omega(k) = ck - \alpha k^3$.

In Fig. 1, curve *3* corresponds to the results of calculating the field on the basis of formula (6). As can be seen from these results, for M lying on the interval $1/2 < M < 2$ the exact field is described by asymptotics (5), whereas for $M > 2$ and with increase in $y/H$, i.e., with increase in the distance from the wave perturbation source, the exact field is described by asymptotics (6).

*Summary*. The asymptotic representations of the solution obtained in the present study make it possible to describe the critical regimes of internal gravity wave generation over a fairly wide velocity range, the field asymptotics expressed in terms of the Airy function being valid for source velocities substantially above critical and at large distances from the perturbation source. Thus, the exact formulas need to be used for obtaining the solution only on a relatively small range of source velocities.